\documentclass[runningheads,a4paper]{llncs}

\usepackage{amssymb}
\usepackage{graphicx}
\usepackage{color}

\usepackage{url}
\urldef{\mailsa}\path|{alfred.hofmann, ursula.barth, ingrid.haas, frank.holzwarth,|
\urldef{\mailsb}\path|anna.kramer, leonie.kunz, christine.reiss, nicole.sator,|
\urldef{\mailsc}\path|erika.siebert-cole, peter.strasser, lncs}@springer.com|    

\begin{document}

\mainmatter  % start of an individual contribution

\title{Improved Query Topic Models via Pseudo-Relevant P\'olya Document Models}

%
% space saving
%

\setlength{\belowcaptionskip}{-10pt}
\setlength{\textfloatsep}{0.4cm}

\author{Ronan Cummins}
\institute{The Computer Lab,\\ University of Cambridge, UK\\
\email{ronan.cummins@cl.cam.ac.uk}}

\maketitle

\begin{abstract}
Query-expansion via pseudo-relevance feedback is a popular method of overcoming the problem of vocabulary mismatch and of increasing average retrieval effectiveness. In this paper, we develop a new method that estimates a \emph{query topic model} from a set of pseudo-relevant documents using a new language modelling framework.

We assume that documents are generated via a mixture of multivariate P\'olya distributions, and we show that by identifying the topical terms in each document, we can appropriately select terms that are likely to belong to the \emph{query topic model}. The results of  experiments on several TREC collections show that the new approach compares favourably to current state-of-the-art expansion methods.

%\keywords{Query Expansion, Language Models, P\'olya Process}
\end{abstract}

\section{Introduction}

Query expansion is an effective technique for overcoming the problem of vocabulary mismatch. In pseudo-relevance feedback (PRF), expansion terms are selected from a set $F$ of top ranked documents of an initial retrieval run using a term-selection algorithm and are added to the initial query in an attempt to improve retrieval. Query expansion via this method has been shown to improve average retrieval effectiveness \cite{lv09:3}. The approach can also be used to suggest possible expansion terms to users, or to build topic models at run-time, where a few initial words provide a seed for the topic. In this paper we focus on the problem of estimating effective query topic models via PRF in a new language modelling framework and provide a number of interesting theoretical insights.

The relevance modelling (RM) approach \cite{lavrenko2001relevance} has been shown to be an effective method for PRF. This approach builds a relevance model $\vec{{\theta}}_R$ from the top $|F|$ documents of an initial retrieval run as follows:
\begin{equation} \label{eq:rm1}
p(t|\vec{{\theta}}_R) = \frac{ \sum_{d \in F} p(t|\vec{{\theta}}_{d}) \cdot p(q|\vec{{\theta}}_{d}) }{ \sum_{d' \in F} p(q|\vec{{\theta}}_{d'})}
\end{equation}
where $\vec{{\theta}}_{d}$ is the \emph{smoothed} document model and $p(q|\vec{{\theta}}_{d})$ is the query likelihood score (document score)\footnote{As it is often assumed that $p(\vec{{\theta}}_{d}|q) \propto p(q|\vec{{\theta}}_{d})$ given a uniform prior over the documents.}. The top-$k$ terms are selected from this relevance model and are linearly interpolated with the original query. One weakness with this formulation is that each document model $\vec{{\theta}}_{d}$ includes a background model, and these are incorporated into the relevance model $\vec{{\theta}}_R$. The general motivation for incorporating a background model is to explain non-topical aspects of documents (e.g. common words and noise), while topical aspects are explained by the unsmoothed document model. We argue that using a model which generates general background terms (noise) during feedback is theoretically anomalous and operationally non-optimal. 

Consequently, in this paper we take a different approach to selecting expansion terms by firstly estimating the likelihood that a candidate term was drawn from the topical part of each of the feedback document models, and subsequently estimating a \emph{query topic model} (QTM) by estimating the probability that the term is topically related to the query. We show that this new approach outperforms the original relevance modelling approach to query expansion and also adheres to a number of recently proposed constraints \cite{clinchant2013theoretical} regarding the term-selection function for PRF. Furthermore, we adopt a recently developed document language model \cite{cummins15} that assumes that documents are generated from a mixture of multivariate P\'olya distributions (aka. the Dirichlet-compound-multinomial). We show that this document model is more effective in the feedback step than using the multinomial language model with a Dirichlet prior. The contribution of this paper is three-fold:

%\begin{itemize}
\begin{list}{$\bullet$}{} 
\item We develop a new method for query expansion using PRF called \emph{query topic modelling} (QTM).

\item We use QTM with a recently developed document language model and show that it adheres to a number of recently developed PRF constraints.  

\item We show that the new method outperforms existing state-of-the-art PRF techniques on a number of TREC collections. 
\end{list}
%\end{itemize}

The remainder of the paper is as follows: Section 2 outlines related work in the area of PRF. Section 3 briefly introduces a recent document language modelling approach before developing a new method of estimating query topic models for use with the aforementioned document model. Section 4 presents an analysis of the new feedback model. Section 5 describes the experimental setup and the results of those experiments. Finally, Section 6 concludes with a discussion.

\section{Related Work}

%Query expansion via PRF has been proposed in information retrieval since the early 1970's. The Rocchio algorithm \cite{rocchiorelevance} for relevance feedback in the vector space model proposed modifying the initial query vector using both relevant and non-relevant document vectors. Much of the research into PRF since can be traced back to these ideas.

In the language modelling framework, there has been a number of initial approaches to building query topic models. The idea of a query model was introduced by Zhai \cite{zhai01:mf} and the simple mixture model (SMM) approach to feedback was developed. The SMM approach aims to extract the topical aspects of the top $|F|$ documents assuming that the same multinomial mixture has generated each document in $F$. By fixing the initial mixture parameter ($\lambda_{smm}$), the topical aspects of the top $|F|$ documents can be estimated using Expectation-Maximisation (EM). Regularised mixture models \cite{Tao:2006} have been developed that aim to eliminate some of the free parameters in the SMM. However, this approach has been shown to be inferior to the SMM \cite{lv09:3}.

Lavrenko \cite{lavrenko2001relevance} developed the idea of building generative relevance models (RM) and this idea was extended to pseudo-relevant documents. It was shown that when these relevance models were interpolated with the initial query model (an approach called RM3 \cite{umass4,lv09:3}), they were highly effective for query expansion. As per Eq.~\ref{eq:rm1}, the RM approach linearly combines the smoothed document models of the top $|F|$ documents. Essentially, the model assumes that short queries and long documents are generated by the same relevance model, and as a consequence the traditional relevance model generates noisy non-topical background words. Furthermore, empirical studies suggest \cite{lv09:3} that different document representations are needed for feedback. They have shown that optimal performance with the RM3 method is achieved when the the document model $\vec{{\theta}}_{d}$ in Eq.~\ref{eq:rm1} remains unsmoothed during feedback. Essentially $p(t|\vec{{\theta}}_{d})$ is estimated using the maximum likelihood of a term occurring in a feedback document.\footnote{The optimal RM3 uses $c(t,d)/|d|$ as $p(t|\vec{{\hat{\theta}}}_{d})$ where $c(t,d)$ is the count of term $t$ in a document of $|d|$ tokens.} Although using an unsmoothed document model in the feedback step is the optimal setting (as is confirmed by our experiments in Section~5), the theoretical anomaly remains (i.e. \emph{why are different document representations needed for retrieval and feedback?}). The optimal RM3 approach is known to select common terms (possibly stopwords) and include them in the expanded query. We show that this is because there is a modelling problem when using the RM approach with query-likelihood for short queries.

A pseudo-relevance based retrieval model using the Dirichlet compound multinomial (DCM) \cite{xu08} (aka. multivariate P\'olya distribution) was reported as outperforming the simple mixture model (SMM). However, in that work the initial document retrieval functions varied and the stronger RM3 baseline was not used. We implement and report a similar term-selection scheme using the Dirichlet-compound-multinomial (PDCM) as a generative model of the top $|F|$ documents as a baseline. As advances in document modelling are likely to yield improvements for principled PRF approaches, we also adopt a recently developed document language model based on the multivariate P\'olya distribution \cite{cummins15}. A detailed comparative study \cite{lv09:3} into PRF approaches reports that both RM3 and SMM achieve comparable performance but that RM3 has more stable parameter settings (i.e. performing consistently well when the background model is removed). More recently, positional pseudo-relevance (PRM) models \cite{lv11} have also been developed which incorporate the proximity of candidate expansion terms to query terms in the feedback documents. We include a positional relevance model baseline (PRM2) in our experiments as a state-of-the-art relevance model that uses term proximity information in the set of feedback documents.

Others \cite{fang06,clinchant2011document,clinchant2013theoretical} have studied the properties of the term-selection scheme in PRF. Many of the useful effects outlined by Clinchant \cite{clinchant2013theoretical} are inherited from studies of constraints for document retrieval \cite{fang04}, while others \cite{clinchant2011document} are explicitly developed for PRF. We perform an analysis of the pseudo-relevance approach developed in this work using the five constraints outlined in \cite{clinchant2013theoretical} ({\bf TF}, {\bf Concavity}, {\bf IDF}, {\bf DF}, and {\bf Document length (DL)} effects) and the one non-redundant constraint \cite{clinchant2011document} (the {\bf document score (DS)} effect).

%The {\bf TF effect} captures the intuition that terms that occur more frequently in set of feedback documents are better candidate expansion terms and should receive a higher weight, while the {\bf Concavity effect} ensures that the increase in weight should decay at higher term-frequencies in the feedback set. The {\bf IDF effect} captures the intuition that rarer terms should be promoted if all else is equal. The {\bf DF effect} states that a term that appears in a greater number of pseudo-relevant documents should receive a higher weight to terms occurring in less pseudo-relevant documents (given that the total  occurrences of the term in the set of pseudo-relevant documents are equal), while the {\bf Document length effect} penalises terms that appear in longer documents in the set $F$. Finally, a {\bf document score constraint} \cite{clinchant2011document} captures the intuition that terms occurrences in high scoring pseudo-relevant documents should receive a higher selection weight than term occurrences in lower scoring pseudo-relevant documents.  Interestingly, if the within document term-frequency aspect of the term-selection scheme is concave, the {\bf document frequency constraint} holds \cite{clinchant2011document}. \footnote{In fact, the concave term-frequency aspect implies the {\bf document frequency constraint} dependent on the {\bf document score constraint}. } 

\section{Document and Query Modelling}

Before developing the new query topic modelling approach, we briefly re-introduce a document language model that we intend to use for modelling the documents in the feedback set $F$.

\subsection{Smoothed P\'olya Urn Document Model}

Recently \cite{cummins15} it has been shown that modelling each document as a mixture of multivariate P\'olya distributions improves the effectiveness of ad hoc retrieval. The model is known to capture word burstiness by modelling the dependencies between recurrences of the same word-type. Furthermore, the model ensures that each document adheres to both the \emph{scope} and \emph{verbosity} hypothesis \cite{robertson1994}. Each document is modelled as follows:

\begin{equation} \label{eq:spud}
\vec{{\alpha}}_{d} =  (1-\omega) \cdot \vec{{\alpha}}_{\tau}  +  \omega \cdot \vec{{\alpha}}_{c}
\end{equation}
where $\vec{{\alpha}}_{d}$, $\vec{{\alpha}}_{\tau}$, and $\vec{{\alpha}}_{c}$ are the document model, topic model,\footnote{For the purposes of this paper, we refer to the unsmoothed model as the \emph{topic model} of the document as it explains words not explained by the general background model.} and background model respectively. The hyper-parameter $\omega$ controls the smoothing and is stable at $\omega = 0.8$. Each of these models are multivariate P\'olya distributions with parameters estimated as follows:

\begin{equation}\label{eq:params}
{\vec{\hat{\alpha}}}_{\tau} = \{{m_d \cdot \frac{c(t,d)}{|d|}} : {t \in d}\}    \hspace{50pt}   {\vec{\hat{\alpha}}}_{c} = \{{m_c \cdot \frac{df_t}{\sum_{t'} df_{t'}}} : {t \in C}\}
\end{equation}
where $m_d$ is the number of word-types (distinct terms) in $d$, $c(t,d)$ is the count of term $t$ in document $d$, $|d|$ is the number of word tokens in $d$, $df_t$ is the document frequency of term $t$ in the collection $C$, and $m_c$ is a background mass parameter that can be estimated via numerical methods (see \cite{cummins15} for details). The scale parameters $m_d$ and $m_c$ can be interpreted as beliefs in the parameters ${c(t,d)}/{|d|}$ and ${df_t}/{\sum_{t'} df_{t'}}$ respectively. 

The query-likelihood approach to ranking documents can be used with these document models whereby one estimates the probability that the query is generated from the expected multinomial drawn from each document model (i.e. $E[\vec{{\alpha}}_{d}]$ is a multinomial).\footnote{For the remainder of the paper when we write $p(q|\vec{\alpha}_d)$, we assume that a point estimate (the expectation) of the multivariate P\'olya is taken.} In this approach to retrieval, queries are generated by the expected multinomial as they are typically short and do not tend to exhibit word burstiness. In line with the original work \cite{cummins15}, we refer to this document language model as the SPUD language model.

\subsection{Query Topic Model (QTM)}

In the relevance model approach to expansion, candidate feedback terms are ranked according to the likelihood of the terms in the relevance model, where the relevance model is estimated as per Eq.~\ref{eq:rm1}. However, this model assumes that all the terms in the document are generated by the relevance model. We assume that documents are generated by both a topical model and a background model, where we first need to estimate the probability that the term seen in the document is topical. Instead, given a set of feedback documents $F$, we then rank terms as follows:

\begin{equation} \label{eq:tm1}
p(\vec{{\theta}}_Q|t) = \frac{ \sum_{d \in F} p(\vec{{\alpha}}_{\tau}|t) \cdot p(q|\vec{{\alpha}}_{d}) }{ \sum_{d' \in F} p(q|\vec{{\alpha}}_{d'})}
\end{equation}
which determines the probability that $t$ was generated by the \emph{query topic model}. While this looks somewhat similar to the relevance model approach (RM) \cite{lavrenko2001relevance} as it uses the query-likelihood document score $p(q|\vec{{\alpha}}_{d})$, it differs in that it uses $p(\vec{{\alpha}}_{\tau}|t)$ instead of $p(t|\vec{{\alpha}}_{d})$ in the numerator. The Bayesian inversion ranks terms by the likelihood of the term being generated by the topical part of the document, and then integrates these probabilities over the top $|F|$ pseudo-relevant documents. Subsequently, the resulting probability $p(\vec{{\theta}}_Q|t)$ will be close to $1.0$ when the term is likely be part of the query topic model, and will be low when the term is unlikely to be part of the query topic model. By assuming a uniform prior over the terms, the parameters of the query topic model $\vec{{\theta}}_Q$ can be found by normalising over the number of feedback terms chosen as follows:

\begin{equation} \label{eq:tm2}
p(t|\vec{{\theta}}_Q) = \frac{ p(\vec{{\theta}}_Q|t) }{ \sum_{t'} p(\vec{{\theta}}_Q|t') }
\end{equation}
As mentioned previously, one of the most prominent approaches to PRF (RM3) interpolates the pseudo-relevance model with the original query $q$. We follow this practise and smooth the query topic model with the original query model as follows: 

\begin{equation} \label{eq:interpolation}
p(t|\vec{{\theta}}_{q'}) =  (1-\pi) \cdot p(t|\vec{{\theta}}_q)  +  \pi \cdot p(t|\vec{{\theta}}_Q)
\end{equation}
where the parameter $\pi$ determines how much mass to assign to the query topic model as compared to the original query model. This interpolation is used in many language modelling approaches to feedback (e.g. RM3 \cite{lv09:3} is recovered by substituting Eq.~\ref{eq:rm1} for $p(t|\vec{{\theta}}_Q)$ above) and has been shown to be stable at $\pi \approx 0.5$. The original query distribution is consistent with the model just presented. The terms in short queries are assumed to have been drawn directly from the \emph{query topic model} and are therefore deemed topical with a probability of $1.0$\footnote{This assumption is likely to valid for short queries. However, for longer queries it is likely that some words are generated by a background model and is worth investigating in future work.} which are subsequently normalised to form $p(t|\vec{{\theta}}_q)$.

Thus far we have outlined a general method to estimate the QTM and therefore any plausible document language modelling approach can be used with it. While we have use the notation $\vec{{\alpha}}$ to denote the multivariate P\'olya, the document models can be replaced with the original multinomial (denoted $\vec{{\theta}}$) with Dirichlet priors. In fact, we will show the results of doing so in Section~5.

\subsection{QTM Using SPUD}

We now outline a specific instantiation of the QTM using the SPUD document model outlined in Section~3.1. Given the SPUD language model (Eq.~\ref{eq:spud}) and its parameters estimates (Eq.~\ref{eq:params}), the probability that the term $t$ was generated from the \emph{topical model} $\vec{{\alpha}}_{\tau}$ of a document can be calculated via Bayes' theorem as follows:

\begin{equation} \label{eq:topical}
p(\vec{{\alpha}}_{\tau}|t) =  \frac{(1-\omega) \cdot {{\alpha}}_{\tau_t}  }{(1-\omega) \cdot {{\alpha}}_{\tau_t}  +  \omega  \cdot {{\alpha}}_{c_t}}
\end{equation}
where ${{\alpha}}_{\tau_t}$ and ${{\alpha}}_{c_t}$ are the parameters of $t$ for the document topic model and background model respectively. A relatively simple intuition for this formula is that topical terms are those that are more likely generated from the topical part of a document than those that are generated by the background model. Interestingly, when plugging in the exact parameters for term $t$, the expression can be re-written in the following form:

\begin{equation} \label{eq:topical2}
p(\vec{{\alpha}}_{\tau}|t) =  \frac{ c(t,d) }{  c(t,d)  +    \frac{\omega \cdot m_c \cdot df_t}{(1- \omega) \cdot \sum_{t'} df_{t'}} \cdot \frac{ |d|}{m_d}}
\end{equation}
where one can notice a concave term-frequency factor not dissimilar to the BM25 term-frequency factor (i.e. $\frac{c(t,d)}{c(t,d) + k_1}$). It should also be remarked that the formula inherits verbosity normalisation from the SPUD model as $|d|/m_d$ is the average term-frequency in the document. We will analyse QTM$_{spud}$ more formally in the next section. For completeness, using the multinomial model with Dirichlet-priors leads to QTM$_{dir}$ as $p(\vec{{\theta}}_{\tau}|t) = \frac{ c(t,d) }{  c(t,d)  +  \mu \cdot p(t|\theta_c)}$ where $p(t|\theta_c)$ is the maximum likelihood of seeing $t$ in the collection $c$.

\section{Analysis}

In this section, we conduct two analyses of the term selection method brought about by the QTM approach outlined in the previous section. For this analysis, we limit ourselves to analysing five term-selection schemes; namely PDCM, SMM, RM3, QTM$_{dir}$, and QTM$_{spud}$. The PDCM approach assumes that the top $|F|$ documents returned for a query have been generated by a DCM and estimates the parameters given the documents in $F$. Terms are then ranked according to their parameter value. SMM \cite{zhai01:mf}, RM3, and QTM have already been discussed. 

\subsection{Constraint Analysis}

The only two approaches that adhere to the {\bf DS} constraint are RM3 and QTM as they use the query-likelihood score to promote terms that appear in documents that are more likely to be relevant (i.e. are highly scored). Neither SMM nor PDCM use the document score in their term selection scheme as they assume that all documents in $F$ are equally relevant\footnote{This is a reasonable assumption for real relevance feedback.}. In the analysis of the remaining constraints, for simplicity we assume that documents in $F$ have received the same document score (are all equally relevant). 

All methods have a term-frequency aspect ({\bf TF}) but this term-frequency aspect is not concave in the case of RM3 (i.e. using maximum likelihood estimates it is easy to see that $c(t,d)/|d|$ is a linear function). PDCM and RM3 do not adhere to the {\bf IDF} constraint as PDCM has no background information and RM3 promotes common terms even when smoothing with a background language model. Most methods penalise the weight contribution from terms in longer documents so {\bf DL} is satisfied for most methods. The only exception is QTM$_{dir}$, this is because the document length is absent in $p(\vec{\theta}_\tau|t)$ due to a cancellation of terms. Finally the {\bf DF} constraint ensures that we should promote terms that appear in more pseudo-relevant documents when all else is equal. Adherence to this constraint follows when the {\bf Concavity} constraint is satisfied \cite{clinchant2011document}.\footnote{Space restricts the complete mathematical formalisms from being presented in this work. The implementation of all approaches are available for download.}

\begin{table*}[!ht] 
\caption{Adherence to Constraints}
\centering
\small
{\renewcommand{\arraystretch}{1.0}
\begin{tabular}{|  l || l | l | l | l | l | l |}

\hline
Method	&			DS &TF	&	Concavity	&	IDF	&	DL	&	DF	\\

\hline															
PDCM			& 	no		&			yes	&	yes			&	no	&	yes	&	yes		\\
SMM				& 	no 		&			yes	&	yes			&	yes	&	yes	&	no		\\
RM3				&	yes 	&			yes	&	no			&	no	&	yes	&	no		\\
QTM$_{dir}$		& 	yes 	& 		yes	&	yes			&	yes	&	no	&	yes		\\
QTM$_{spud}$	& 	yes 	& 		yes	&	yes			&	yes	&	yes	&	yes		\\
\hline

\end{tabular}}
\end{table*}

\subsection{Qualitative analysis}

\begin{table*}[!ht] 
\caption{Top 15 expansion words and their unnormalised term-selection value according to four PRF approaches. In all approaches the initial retrieval method is the SPUD language model with $\omega=0.8$ and the set of pseudo relevant documents $|F|=10$. Terms in red are those that receive a score of less than $0.5$ according to the QTM$_{spud}$ model}
\centering

\scriptsize
%\begin{tabular}{| p{3cm} | p{3cm} | p{3cm} | p{3cm}| p{3cm}|}
{\renewcommand{\arraystretch}{1.1}
\begin{tabular}{|  p{1.3cm} | p{1.5cm} | p{0.8cm}| p{1.5cm}| p{0.8cm} | p{1.5cm} | p{0.8cm} | p{1.5cm}| p{0.8cm}|}
\hline
 & 
\multicolumn{8}{c|}{PRF methods for Topic 697 in {\bf robust-04} } \\

\hline
Query & 
\multicolumn{8}{c|}{{\bf \emph{air traffic control}}} \\
\hline

Method & 
\multicolumn{2}{c|}{PDCM}	& 
\multicolumn{2}{c|}{SMM$_{\lambda=0.2}$}	& 
\multicolumn{2}{c|}{RM3$_{\omega=0}$}	& 
\multicolumn{2}{c|}{QTM$_{spud}$}\\

\hline
1	&	{\bf air}	&	71.159	&	{\bf air	}&	0.0793	&	{\bf control}	&	0.0313	&	{\bf traffic	}&	0.9835	\\
2	&	{\bf control	}&	68.542	&	{\bf control	}&	0.0749	&	{\bf air}	&	0.0310	&	{\bf air}	&	0.9619	\\
3	&	{\bf traffic}	&	56.838	&	{\bf traffic}	&	0.0655	&	{\bf traffic}	&	0.0250	&	{\bf control}	&	0.9227	\\
4	&	system	&	33.123	&	system	&	0.0350	&	system	&	0.0125	&	aviat	&	0.8795	\\
5	&	year	&	25.052	&	\textcolor{red}{atc}	&	0.0216	&	year	&	0.0109	&	airlin	&	0.8668	\\
6	&	\textcolor{red}{said}	&	21.862	&	airport	&	0.0149	&	\textcolor{red}{said}	&	0.0105	&	airport	&	0.8389	\\
7	&	\textcolor{red}{from}	&	16.890	&	safeti	&	0.0137	&	\textcolor{red}{new}	&	0.0072	&	transport	&	0.7684	\\
8	&	problem	&	15.871	&	aviat	&	0.0135	&	\textcolor{red}{from}	&	0.0071	&	flight	&	0.7319	\\
9	&	\textcolor{red}{new}	&	15.195	&	airlin	&	0.0134	&	\textcolor{red}{european}	&	0.0070	&	system	&	0.7141	\\
10	&	\textcolor{red}{ha}	&	13.754	&	\textcolor{red}{faa}	&	0.0128	&	problem	&	0.0067	&	safeti	&	0.6251	\\
11	&	airport	&	13.625	&	flight	&	0.0128	&	airlin	&	0.0059	&	problem	&	0.6243	\\
12	&	\textcolor{red}{which}	&	13.521	&	problem	&	0.0126	&	\textcolor{red}{ha}	&	0.0056	&	radar	&	0.6196	\\
13	&	\textcolor{red}{have}	&	13.409	&	\textcolor{red}{european}	&	0.0111	&	safeti	&	0.0055	&	inadequ	&	0.6132	\\
14	&	safeti	&	12.724	&	\textcolor{red}{facil}	&	0.0103	&	airport	&	0.0054	&	rout	&	0.5859	\\
15	&	airlin	&	12.695	&	\textcolor{red}{europ}	&	0.0100	&	\textcolor{red}{europ}	&	0.0053	&	delai	&	0.5552	\\
%16	&	\textcolor{red}{us}	&	11.879	&	airspac	&	0.0097	&	\textcolor{red}{have}	&	0.0053	&	improv	&	0.5516	\\
%17	&	\textcolor{red}{european}	&	11.846	&	\textcolor{red}{tracon}	&	0.0078	&	\textcolor{red}{which}	&	0.0050	&	jet	&	0.5476	\\
%18	&	aviat	&	11.798	&	\textcolor{red}{moor}	&	0.0075	&	\textcolor{red}{would}	&	0.0048	&	year	&	0.5440	\\
%19	&	\textcolor{red}{than}	&	11.535	&	transport	&	0.0074	&	aviat	&	0.0046	&	facil	&	0.5427	\\
%20	&	\textcolor{red}{atc}	&	11.384	&	delai	&	0.0070	&	flight	&	0.0045	&	workload	&	0.5271	\\
\hline

\end{tabular}}
\label{fig:example_terms}
\end{table*}

Fig.\ref{fig:example_terms} shows the top 20 terms selected from four PRF approaches. QTM$_{dir}$ (not shown) returns term very similar to those returned by QTM$_{spud}$. The score for each of the terms is in its unnormalised form. We see that the two methods that do not adhere to the {\bf IDF} constraint (PDCM and RM3) tend to select high frequency words in the top $|F|$ documents without regard to their distribution in the entire collection. Although these frequent terms might not be highly detrimental when added to the initial query, it suggests that more expansion terms may be needed in order to achieve optimal performance. From a qualitative perspective, the QTM approach appears to promote expansion terms that are more semantically coherent when compared to PDCM and RM3. This would be of use in applications where one wished to generate topic models given a few initial terms. Furthermore, we can see that the score of the QTM$_{spud}$ approach has an intuitive interpretation as the probability that the term belongs to the query topic model. All of the terms in red are those that are more likely to have been generated by the background model according to QTM$_{spud}$.\footnote{It would be interesting future work to investigate only selecting terms above a certain threshold (e.g. those terms that are more likely than not to be topical i.e. $p(\vec{\alpha_{\tau}}|t)>0.5$).}

\section{Experimental Evaluation}

%\subsection{Experimental Design}

Our experiments have three main aims. Firstly we wish to determine if QTM is empirically consistent with its theoretical derivation. Therefore, we aim to show that during feedback the smoothed document models are effective and stable when using a similar parameter to that used during the initial retrieval step. Secondly, we aim to determine the effectiveness of the new QTM model for query expansion when compared with a number of \emph{state-of-the-art} approaches. Finally, we aim to validate our choice of document model (multinomial vs multivariate P\'olya) in the feedback step. 

To these ends, we used a number of standard TREC\footnote{\url{http://trec.nist.gov/}} collections (robust-04, wt2g, wt10g, gov2, and ohsumed). Stemming and stopword removal (a small list of less than 30 words) was performed. The title fields of the associated topics are used as queries. As a weak baseline we use a tuned language model with Dirichlet priors for retrieval (Dir$_{\hat{\mu}}$) and use the RM3 approach with $\mu=0$ during the feedback step. This is currently a strong operational baseline. As a stronger set of baselines we use the SPUD$_{\omega=0.8}$ approach for retrieval with feedback approaches of PDCM, the simple mixture model (SMM), and the relevance model (RM3). Finally, we used a reportedly stronger positional relevance model baseline (PRM2) \cite{lv11} that uses proximity information in the feedback documents where we set the proximity parameter to its suggested value $\sigma=200$ \cite{lv11}. In all experiments document retrieval is performed using the same function for both the original query and the expanded query. 

To ensure a fair comparison, terms are ranked according to the selection function for each approach, are then normalised to sum to $1.0$, and interpolated with the original query using $\pi$ in Eq.~\ref{eq:interpolation}. We tuned the three parameters $\pi$, $|F|$, and the number of feedback terms $|T|$ using two-fold cross-validation\footnote{using even and odd numbered topics as our two folds.}. All approaches were implemented in Lucene and the code needed to replicate all of the results in this paper is available for download.\footnote{\url{https://github.com/ronancummins/query-topic-model}}

\subsection{Results}

\begin{figure}[!ht] 
\begin{center}
\[\arraycolsep=0.0pt\def\arraystretch{0.0}
\begin{array}{c c c}
   	{\includegraphics[height=2.0cm,width=3.0cm]{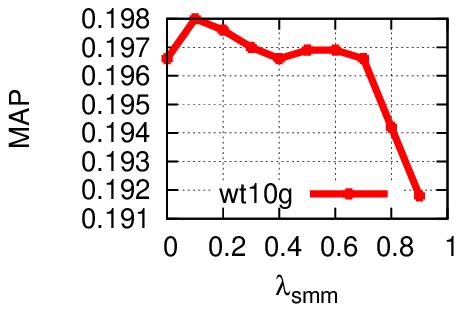}} &    
	\includegraphics[height=2.0cm,width=3.0cm]{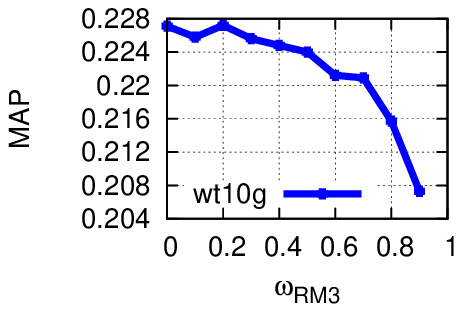} & 
	\includegraphics[height=2.0cm,width=3.0cm]{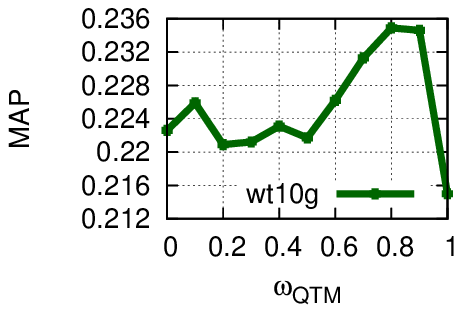} \\
   	\includegraphics[height=2.0cm,width=3.0cm]{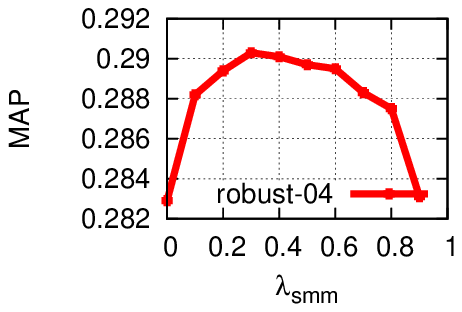} &    
	\includegraphics[height=2.0cm,width=3.0cm]{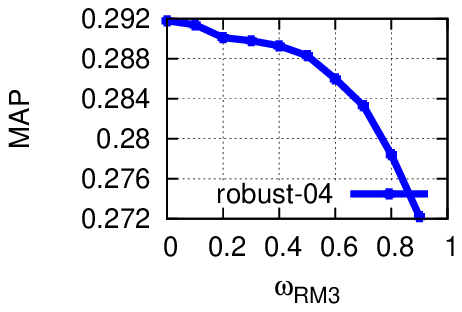} &
	\includegraphics[height=2.0cm,width=3.0cm]{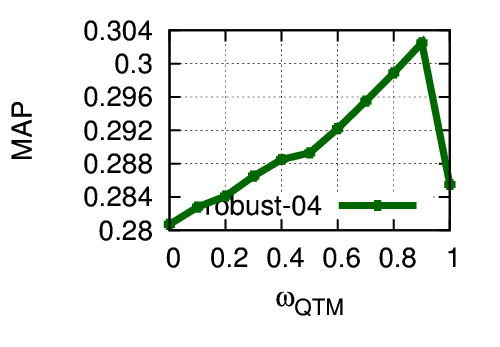} \\
   	{\includegraphics[height=2.0cm,width=3.0cm]{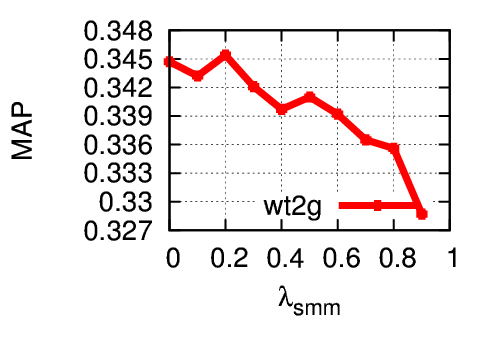}} &    
	\includegraphics[height=2.0cm,width=3.0cm]{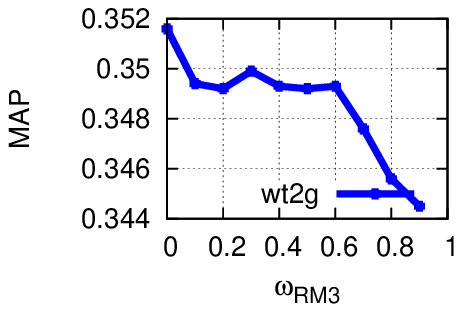} & 
	\includegraphics[height=2.0cm,width=3.0cm]{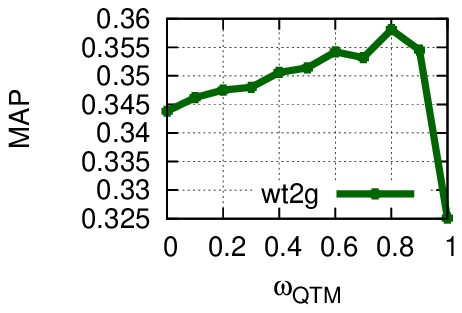} 	
\end{array}
\]
\caption{Retrieval effectiveness as smoothing parameter in feedback step changes in three PRF approach (SMM, RM3, and QTM from left to right).}
\label{fig:smoothing}
\end{center}
\end{figure}

Fig.~\ref{fig:smoothing} shows the effectiveness of three PRF approaches (SSM, RM3, and QTM$_{spud}$) as the background mass changes on three TREC collections (PDCM does not use a background model) during feedback. The same retrieval method (SPUD) was used in this experiment. The SMM approach is relatively stable on these test collections at $\lambda_{smm}=0.2$. We can see that the RM3 approach is most effective when using no smoothing ($\omega = 0.0$). This is consistent with previous research using the multinomial with Dirichlet priors \cite{lv09:3} and confirms that different document representations are needed for initial retrieval and feedback when using RM3. The QTM$_{spud}$ approach is most effective using the same background mass parameter that is used in the initial retrieval (i.e. $\omega = 0.8$). This result confirms that the background language model has useful information for term-selection. This also suggests that the QTM model is theoretically more consistent than RM3 as the same document representation is appropriate for initial retrieval and feedback. The remaining collections (ohsumed and gov2, not included in Fig.~\ref{fig:smoothing} due to space restrictions) show the same trend.

\begin{table*}[!ht] 
\caption{MAP (NDCG@10) of PRF approaches on 5 test collections ($*$ means statistically significant compared to SPUD-RM3$_{\omega=0}$ at $p<0.05$ using a paired t-test, while $\dagger$ means statistically significant when compared with QTM$_{dir}$ at $p<0.05$. The best result per collection is in {\bf bold}).} \label{tab:res}
\centering

\scriptsize
\setlength{\tabcolsep}{1pt}
{\renewcommand{\arraystretch}{1.2}
\begin{tabular}{|  c | l || l | l | l | c | c | }

\hline
	&				&	ohsu	&	robust-04	&	wt2g	&	wt10g	&	gov2	\\
		&	\# docs			&	283k docs	&	528k	&	247k	&	1.69M	&	25.2M	\\
		&	topics			&	1-63	 &	301-450,	&	401-500	&	450-550	&	701-850	\\
		&					&		 &	601-700	&		&			&		\\
		&	\# queries			&	63	&	249	&	50	&	100	&	149	\\
\hline															
Retrieval	&	Expansion			&	\multicolumn{5}{c|}{}		\\
%Function	&	Function			&	\multicolumn{5}{c|}{}		\\
\hline															
Dir$_{\hat{\mu}}$	&	None			&	0.321 (0.516)	&	0.256 (0.466)	&	0.311 (0.490)	&	0.194 (0.347)	&	0.303 (0.573)	\\
Dir$_{\hat{\mu}}$	&	RM3$_{\mu=0}$			&	0.374 (0.564)	&	0.288 (0.484)	&	0.346 (0.514)	&	0.213 (0.353)	&	0.332 (0.575)	\\
\hline															
SPUD	&	None			&	0.327 (0.520)	&	0.260 (0.480)	&	0.316 (0.495)	&	0.204 (0.366)	&	0.315 (0.596)	\\
SPUD	&	SMM$_{\lambda=0.2}$			&	0.375 (0.568)	&	0.285 (0.471)	&	0.334 (0.510)	&	0.212 (0.363)	&	0.329 (0.568)	\\
SPUD	&	PDCM			&	0.376 (0.565)	&	0.293 (0.489)	&	0.340 (0.511)	&	0.213 (0.368)	&	0.338 (0.598)	\\
SPUD	&	PRM2			&	0.379 (0.567)	&	{\bf 0.305} ({\bf 0.496})	&	0.359 ({\bf 0.539})	&	0.225 ({\bf 0.371})	&	{\bf 0.350} (0.609)	\\
SPUD	&	RM3$_{\omega=0}$		&	0.374 (0.572)	&	0.302 (0.494)	&	0.355 (0.535)	&	0.216 (0.362)	&	 0.348 (0.604)	\\
\hline															
SPUD	&	QTM$_{dir}$			&	0.380 (0.558)	&	0.297 (0.491)	&	0.357 (0.517)	&	0.217 (0.357)	&	0.345 (0.628)	\\
SPUD	&	QTM$_{spud}$			&	{\bf 0.384$*$} ({\bf 0.579})	&	{0.300$\dagger$} ({0.493})	&	{\bf 0.364$\dagger$} (0.529)	&	{\bf 0.220$*$} (0.374)	&	0.345 ({\bf 0.632$*$})	\\
\hline

\end{tabular}}
\end{table*}

Table~\ref{tab:res} shows the effectiveness (MAP and NDCG@10) of the QTM model compared to the baselines on five test collections. The QTM$_{spud}$ approach significantly outperforms the tuned RM3 approach on a number of collections. It is surprising that QTM$_{spud}$ is competitive with the positional relevance model (PRM2) which uses proximity information. Furthermore, the QTM$_{spud}$ approach outperforms the QTM$_{dir}$ approach confirming that the P\'olya document models are also better than the multinomial document models for feedback. This also suggests that the {\bf DL} constraint is advantageous as it is the main difference between these methods. The improvements of QTM$_{spud}$ over QTM$_{dir}$ are consistent but small in magnitude. 

\begin{figure}[!ht]
\begin{center}
\[\arraycolsep=0.0pt\def\arraystretch{0.0}
\begin{array}{c c c}
   	{\includegraphics[height=2.5cm,width=3.5cm]{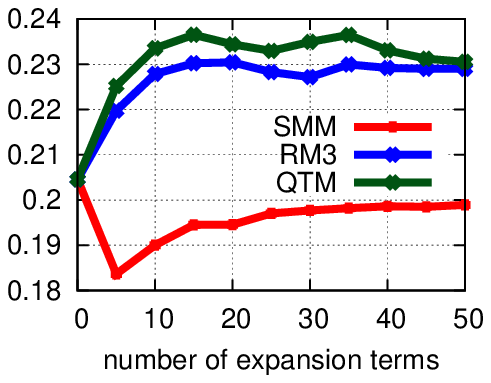}} &    
	\includegraphics[height=2.5cm,width=3.5cm]{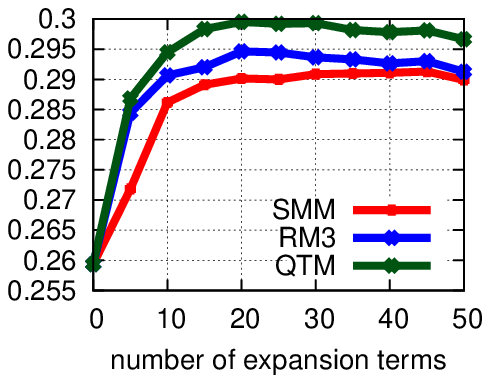} & 
	\includegraphics[height=2.5cm,width=3.5cm]{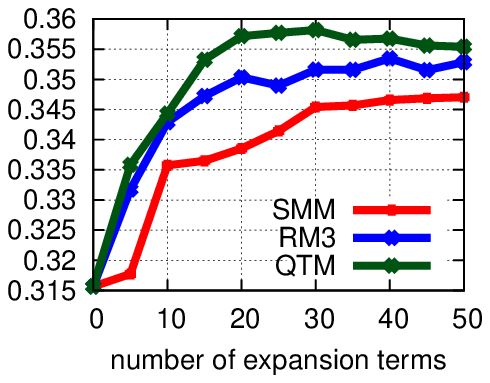} 
\end{array}
\]
\caption{Retrieval effectiveness as number of expansion terms increase for three PRF approaches (SMM, RM3, and QTM$_{spud}$) on three collections (wt10g, robust-04, and wt2g from left to right).}
\label{fig:terms}
\end{center}
\end{figure}

Finally, Fig.~\ref{fig:terms} shows the performance of three approaches when the number of expansion terms vary. SMM is the worst approach and QTM$_{spud}$ outperforms RM3. These differences tend to be less pronounced as more terms are added. We hypothesise that this is because as the number of expansion terms increase, the same terms tend to get added to the initial query. However, QTM$_{spud}$ retains its performance advantage when adding fewer expansion terms. In fact, during cross-validation we found that the optimal number of expansion terms for QTM is lower than for any of the other expansion methods studied here.

\section{Discussion and Conclusion}
The QTM approach developed in this work is similar in spirit to the simple mixture model (SMM) outlined in the original work of Zhai and Lafferty \cite{zhai01:mf}. However, there is no closed-form solution for the SMM approach and there is a free-parameter for which there is no obvious way of determining a suitable value (aside from tuning it empirically). While RM3 has stable performance, it is when different document representations are used for feedback (i.e. no background mass). Conversely for the QTM approach, we have shown that the same hyper-parameter values used to smooth documents for retrieval (i.e. $\omega=0.8$ for SPUD), are close to optimal during the feedback process as shown in Fig.~\ref{fig:smoothing}. This, unlike RM3, gives theoretical consistency to our approach. QTM achieves good performance at $|F|=10$, $\pi=0.5$, and with $30$ expansion terms.

A brief analysis of the QTM approach has shown that it adheres to a number of previously proposed properties describing effective term-selection functions. It is interesting that these properties arise from our approach without manipulating or heuristically hand-crafting the function in any way. A qualitative analysis of the terms selected by the QTM$_{spud}$ indicates they are more topically coherent than those selected by RM3. This is because at its most optimal setting, RM3 selects the most frequent terms in the feedback documents without regard to their distribution in the collection. The QTM approach is competitive with several strong baselines, including a positional relevance model, when using the same retrieval method. Future work will look at developing better expansion models for use with verbose queries.

\bibliographystyle{plain}
\bibliography{query_expansion_v2}

\end{document}